\documentclass[twocolumn,reprint,showpacs,amssymb,amsmath,aps,pra]{revtex4-1}
\newcommand{\half}{\mbox{$\frac{1}{2}$}}
\usepackage{graphics}
\usepackage{bm}
\usepackage{graphics}
\usepackage{bm}
\usepackage{amsmath}
\usepackage{amssymb}
\begin{document}

\title{Generalized entropic measures of quantum correlations}
\author{R. Rossignoli, N. Canosa, L. Ciliberti}
\affiliation{Departamento de F\'{\i}sica-IFLP,
Universidad Nacional de La Plata, C.C. 67, La Plata (1900) Argentina}
\begin{abstract}
We propose a general measure of non-classical correlations for bipartite
systems based on generalized entropic functions and majorization properties.
Defined as the minimum information loss due to a local measurement, in the case
of pure states it reduces to the generalized entanglement entropy, i.e., the
generalized entropy of the reduced state. However, in the case of mixed states
it can be non-zero in separable states, vanishing just for states diagonal in a
general product basis, like the Quantum Discord. Simple quadratic measures of
quantum correlations arise as a particular case of the  present formalism. The
minimum information loss due to a joint local measurement is also discussed.
The evaluation of these measures in a few simple relevant cases is as well
provided, together with comparison with the corresponding entanglement
monotones.

\end{abstract}
\maketitle
\section{Introduction}
Quantum entanglement is well known to be an essential resource for performing
certain quantum information processing tasks such as quantum teleportation
\cite{Be.93,NC.00}. It has also been shown to be essential for achieving an
exponential speed-up over classical computation in the case of pure-state based
quantum computation \cite{JL.03}. However, in the case of mixed-state quantum
computation, such as the model of Knill and Laflamme \cite{KL.98}, such
speed-up can be achieved without a substantial presence of entanglement
\cite{DFC.05}. This fact has turned the attention to other types and measures
of quantum correlations, like the quantum discord (QD) \cite{OZ.01,HV.01},
which, while reducing to the entanglement entropy in bipartite pure states, can
be non-zero in certain separable mixed states involving mixtures of
non-commuting product states. It was in fact shown in \cite{DSC.08} that the
circuit of \cite{KL.98} does exhibit a non-negligible value of the QD between
the control qubit and the remaining qubits. As a result, interest on the QD
\cite{LBAW.08,Lu.08,Sa.09,SL.09,FA.10} and other alternative measures of
quantum correlations for mixed states \cite{Lu2.08,DG.09,WPM.09,Ve.10} has
grown considerably.

The aim of this work is to embed measures of quantum correlations within a
general formulation based on majorization concepts \cite{NC.00,Wh.78,Bh.97} and
the generalized information loss induced by a measurement with unknown result.
This framework is able to provide general entropic measures of quantum
correlations for mixed quantum states with properties similar to those of the
QD, like vanishing just for states diagonal in a standard or conditional
product basis (i.e., classical or partially classical states) and reducing to
the corresponding generalized entanglement entropy in the case of pure states.
But as opposed to the QD and other related measures \cite{Lu2.08,Ve.10}, which
are based essentially on the von Neumann entropy
\begin{equation}
S(\rho)=-{\rm Tr}\,\rho\,\log_2 \rho\,,\label{S}
\end{equation}
and rely on specific associated properties, the present measures are applicable
with general entropic forms satisfying minimum requirements \cite{Wh.78,CR.02}.
For instance, they can be directly applied with the linear entropy
\begin{equation}
S_2(\rho)=2(1-{\rm Tr}\,\rho^2)\,,\label{SL}
\end{equation}
which corresponds to the linear approximation $-\ln\rho\approx 1-\rho$ in
(\ref{S}) and is directly related to the purity ${\rm Tr}\,\rho^2$ and the pure
state concurrence \cite{Ca.03,WW.98}, and whose evaluation in a general
situation is easier than (\ref{S}) as it does not require explicit knowledge of
the eigenvalues of $\rho$. We will show, however, that the same qualitative
information can nonetheless be obtained. The positivity of the QD relies on the
special concavity property of the conditional von Neumann entropy
\cite{OZ.01,Wh.78,NC.00}, which prevents its direct extension to general
entropic forms.

The concepts of generalized entropies, generalized information loss by
measurement and the ensuing entropic measures of quantum correlations based on
minimum information loss due to local or joint local measurements are defined
and discussed in \ref{II}. Their explicit evaluation in three specific examples
is provided in \ref{III}, where comparison with the corresponding entanglement
monotones is also discussed. Conclusions are finally drawn in \ref{IIII}.

\section{Formalism\label{II}}
\subsection{Generalized entropies}
Given a density operator $\rho$ describing the state of a quantum system
($\rho\geq 0$, ${\rm Tr}\,\rho=1$), we define the generalized entropies
\cite{CR.02}
\begin{equation}
S_f(\rho)={\rm Tr}\,f(\rho)\,, \label{Sf}
\end{equation}
where $f(p)$ is a smooth strictly concave real function defined for $p\in[0,1]$
satisfying $f(0)=f(1)=0$ ($f$ is continuous in $[0,1]$ and $f'$ strictly
decreasing in $(0,1)$, such that $f(qp_i+(1-q)p_j)>qf(p_i)+(1-q)f(p_j)$
$\forall$ $q\in(0,1)$ and $p_i\neq p_j$). We will further assume here
$f''(p)<0$ $\forall$ $p\in(0,1)$, which ensures strict concavity. As in
(\ref{S})--(\ref{SL}), we will normalize entropies such that $S_f(\rho)=1$ for
a maximally mixed single qubit state ($2f(1/2)=1$). While our whole discussion
can be directly extended to more general concave or Schur-concave \cite{Bh.97}
functions, we will concentrate here on the simple forms (\ref{Sf}) which
already include many well known instances: The von Neumann entropy (\ref{S})
corresponds to $f(p)=-p\log_2 p$, the linear entropy  (\ref{SL}) to
$f(p)=2(p-p^2)$, and the Tsallis entropy \cite{Ts.09} $S_q(\rho)\propto 1-{\rm
Tr}\,\rho^q$ to $f(p)=(p-p^q)/(1-2^{1-q})$ for the present normalization, which
is concave for $q>0$. It reduces to the linear entropy (\ref{SL}) for $q=2$ and
to the von Neumann entropy (\ref{S}) for $q\rightarrow 1$. The R\'{e}nyi
entropy \cite{Wh.78} $S_q^R(\rho)=(\log_2\,{\rm Tr}\,\rho^q)/(1-q)$ is just an
increasing function of $S_q(\rho)$. The Tsallis entropy has been recently
employed to derive generalized monogamy inequalities \cite{Ki.10}. Entropies of
the general form (\ref{Sf}) were used to formulate a generalized entropic
criterion for separability \cite{RC.02,RC.03}, on the basis of the majorization
based disorder criterion \cite{NK.01}, extending the standard entropic criterion \cite{HH.96}.

While additivity amongst the forms (\ref{Sf}) holds only in the von Neumann
case ($S(\rho_A\otimes \rho_B)=S(\rho_A)+S(\rho_B)$), strict concavity and the
condition $f(0)=f(1)=0$ ensure that all entropies (\ref{Sf}) satisfy
\cite{CR.02}: i) $S_f(\rho)\geq 0$, with $S_f(\rho)=0$ if and only if (iff)
$\rho$ is a pure state ($\rho^2=\rho$), ii) they are concave functions of
$\rho$ ($S_f(\sum_i q_i\rho_i)\geq \sum_i q_i S_f(\rho_i)$ if $q_i\geq 0$,
$\sum_i q_i=1$) and iii) {\it they increase with increasing mixedness}
\cite{Wh.78}:
\begin{equation}
\rho'\prec\rho \Rightarrow S_f(\rho')\geq S_f(\rho)\,,\label{mf}
 \end{equation}
where $\rho'\prec\rho$ indicates that $\rho'$ is {\it majorized} by $\rho$
\cite{Bh.97,Wh.78}:
\begin{equation}
\rho'\prec\rho\Leftrightarrow \sum_{j=1}^i p'_j\leq \sum_{j=1}^i\,
p_j,\;\;i=1,\ldots n-1\,. \label{m2}
 \end{equation}
Here $p_i$, $p'_i$ denote the eigenvalues of $\rho$ and $\rho'$ sorted in {\it
decreasing} order ($p_i\geq p_{i+1}$, $\sum_{i=1}^n p_i=1$) and $n=n'$ the
dimension of $\rho$ and $\rho'$ (if different, the smaller set of eigenvalues
is to be completed with zeros). Essentially $\rho'\prec\rho$ indicates that the
probabilities $\{p'_i\}$ are more spread out than $\{p_i\}$. The maximally
mixed state $\rho_n=I_n/n$ satisfies $\rho_n\prec\rho$ $\forall$ $\rho$ of
dimension $n$, implying that all entropies $S_f(\rho)$ attain their maximum at
such state: $S_f(\rho)\leq S_f(\rho_n)=n\,f(1/n)$ $\forall$ $\rho$ of rank
$r\leq n$.

Eq.\ (\ref{mf}) follows from concavity (and the condition $f(0)=0$ if $n\neq
n'$) as for $n=n'$, $\rho'\prec\rho$ iff $\rho'$ is a mixture unitaries of
$\rho$ \cite{Wh.78,NC.00} ($\rho'=\sum_i q_i U_i\rho U^\dagger_i$, $q_i>0$,
$U_i^\dagger U_i=I$), and $S_f(U_i\rho U_i^\dagger)=S_f(\rho)$. Moreover, if at
least one of the inequalities in (\ref{m2}) is strict ($<$), then
$S_f(\rho')>S_f(\rho)$, as $S_f(\rho)=\sum_{i}f(p_i)$ is a strictly decreasing
function of the partial sums $s_i=\sum_{j=1}^i p_j$ \cite{RC.02} ($\partial
S_f/\partial s_i=f'(p_i)-f'(p_{i+1})<0$ if $p_{i+1}<p_i$, $i<n$).

While the converse of Eq.\ (\ref{mf}) does not hold in general ($S_f(\rho')\geq
S_f(\rho)\Rightarrow\!\!\!\!\!\!\!/ \;\rho'\prec\rho$), it does hold if valid
for {\it all} $S_f$ of the present form (an example of a smooth sufficient set
was provided in \cite{RC.03}):
\begin{equation}
S_f(\rho')\geq S_f(\rho)\;\forall\;S_f\Rightarrow \rho'\prec\rho\,.\label{Sfp}
\end{equation}
Hence, although the rigorous concept of disorder implied by majorization
($\rho'\prec\rho$) cannot be captured by any single choice of entropy,
consideration of the general forms (\ref{Sf}) warrants complete correspondence
through Eq.\ (\ref{Sfp}).

\subsection{Generalized information loss by measurement}
Let us now consider a general projective measurement $M$ on the system,
described by a set of orthogonal projectors $P_k$ ($\sum_k P_k=I_n$, $P_k
P_{k'}=\delta_{kk'}P_k$). The state of the system after this measurement, if
the result is unknown, is given by \cite{NC.00}
\begin{equation}\rho'=\sum_k P_k \rho P_k\,,\label{rhop}\end{equation}
which is just the ``diagonal'' of $\rho$ in a particular basis ($\rho'=\sum_j
\langle j'|\rho|j'\rangle|j'\rangle\langle j'|$, with $|j'\rangle$ the
eigenvectors of the blocks $P_k\rho P_k$). It is well known that such diagonals
are always more mixed than the original $\rho$ \cite{Wh.78,Bh.97}, i.e.,
$\rho'\prec\rho$, and hence, for any $f$ of the present form,
\begin{equation}
S_f(\rho')\geq S_f(\rho)\,.\label{Sm}
\end{equation}
Moreover, $S_f(\rho')=S_f(\rho)$ iff $\rho'=\rho$, i.e., if $\rho$ is unchanged
by such measurement (if $\rho=\sum_i p_i|i\rangle\langle i| \neq \rho'$, strict
concavity implies $S_f(\rho')=\sum_j f(\sum_i p_i|\langle
j'|i\rangle|^2)>\sum_{i,j}|\langle j'|i\rangle|^2f(p_i)=\sum_i f(p_i)$). A
measurement with unknown result entails then no gain and most probably a loss
of information according to any $S_f$. The difference
\begin{equation}
I_f^M(\rho)=S_f(\rho')-S_f(\rho)\label{DfM}
 \end{equation}
quantifies, according to the measure $S_f$, this loss of information, i.e., the
information contained in the off-diagonal elements of $\rho$ in the basis
$\{|j'\rangle\}$. It then satisfies $I_f^M(\rho)\geq 0$, with $I_f^M(\rho)=0$
iff $\rho'=\rho$.

In the case of the von Neumann entropy (\ref{S}), Eq.\ (\ref{DfM}) reduces to
the {\it relative} entropy \cite{Wh.78,NC.00,Ve.02} between $\rho$ and $\rho'$,
since their diagonal elements in the basis $\{|j'\rangle\}$ coincide:
\begin{subequations}
\label{IS}
\begin{eqnarray}
I^M(\rho)&=&S(\rho')-S(\rho)\label{IM}\\
&=&{\rm Tr}\,\rho\,(\log_2\rho-\log_2\rho')=S(\rho||\rho')\,. \label{DsB}
\end{eqnarray}
\end{subequations}
The relative entropy $S(\rho||\rho')$ is well known to be non-negative
$\forall$ $\rho,\rho'$, vanishing just if $\rho=\rho'$ \cite{Wh.78,NC.00}. In
the case of the linear entropy (\ref{SL}), Eq.\ (\ref{DfM})  becomes instead
\begin{subequations}
\label{I2}
\begin{eqnarray}
I_2^{M}(\rho)&=&2{\rm Tr}(\rho^2-{\rho'}^2)\label{I2M}\\
&=&2{\rm Tr}\,\rho(\rho-\rho')=2||\rho-\rho'||^2\label{D2b}\,,
\end{eqnarray}
\end{subequations}
where $||A||=\sqrt{{\rm Tr}A^\dagger A}$ is the Hilbert-Schmidt or Frobenius
norm. Hence, $I_2^M(\rho)$ is just the square of the norm of the off-diagonal
elements in the measured basis, being again verified that $I_2^M(\rho)=0$ only
if $\rho'=\rho$.

Let us remark, however, that the general positivity of (\ref{DfM}) arises just
from the majorization $\rho'\prec\rho$ and the strict concavity of $S_f$, the
specific properties of the measures (\ref{DsB})--(\ref{D2b}) being not invoked.
In fact, if the off-diagonal elements of $\rho$ in the measured  basis are
sufficiently small, a standard perturbative expansion of (\ref{DfM}) shows that
\begin{equation}
I_f^M(\rho)\approx \sum_{j<k}\frac{f'(p'_{k})-f'(p'_j)}{p'_j-p'_k}
|\langle j'| \rho|k'\rangle|^2\,,\label{quad}
 \end{equation}
where $p'_j=\langle j'|\rho|j'\rangle$. The fraction in (\ref{quad}) is
positive $\forall$ $p'_j\neq p'_k$ due to the concavity of $f$ (if $p'_j=p'_k$,
it should be replaced by $-f''(p'_j)<0$). Eq.\ (\ref{quad}) is just the
square of a {\it weighted} quadratic norm of the off-diagonal elements. In the
case (\ref{SL}), Eq.\ (\ref{quad}) reduces of course to Eq.\ (\ref{D2b}).

For generalized measurements \cite{NC.00} leading to
\begin{equation}
\rho'=\sum_k M_k \rho M_k^\dagger\,,\label{Mk}
\end{equation}
Eq.\ (\ref{Sm}) and the positivity of (\ref{DfM}) remain valid $\forall$ $S_f$
if {\it both} conditions i) $\sum_k M_k^\dagger M_k=I$ and ii) $\sum_k M_k
M_k^\dagger=I$ are fulfilled: if $|j'\rangle$ and $|i\rangle$ denote the
eigenvectors of $\rho'$ and $\rho$, we then have $\sum_{j,k}|\langle
j'|M_k|i\rangle|^2= \sum_{i,k}|\langle j'|M_k|i\rangle|^2=1$ and hence
$S_f(\rho')=\sum_j f(\sum_{k,i}|\langle j'|M_k|i\rangle|^2p_i)\geq
\sum_{j,k,i}\!\!|\langle j'|M_k|i\rangle|^2f(p_i)=\!\sum_i f(p_i)$, i.e.,
$\rho'\prec\rho$. While i) ensures trace conservation, ii) warrants that the
eigenvalues of $\rho'$ are convex combinations of those of $\rho$. If not
valid, Eq.\ (\ref{Sm}) no longer holds in general, as already seen in trivial
single qubit examples ($M_0=|0\rangle\langle 0|$, $M_1=|0\rangle \langle 1|$
will change any state $\rho$ into the pure state $|0\rangle\langle 0|$, yet
fulfilling i)). For projective measurements, $M_k=P_k$.

\subsection{Minimum information loss by a local measurement}
Let us now consider a bipartite system $A+B$ whose state is specified by a
density matrix $\rho_{AB}$. Suppose that a complete {\it local} measurement
$M_B$ in system $B$ is performed, defined by one dimensional local projectors
$P^B_j=|j_B\rangle\langle j_B|$. The state after this measurement (Eq.\
(\ref{rhop}) with $P_k\rightarrow I_A\otimes P_j^B$) becomes
\begin{equation}
\rho'_{AB}=\sum_j q_j\, \rho_{A/j}\otimes P^B_j\,,\label{rhopab}
 \end{equation}
where $q_j={\rm Tr}[\rho_{AB} I_A\otimes P^B_j]$ is the probability of outcome
$j$ and $\rho_{A/j}={\rm Tr}_B [\rho_{AB} I_A\otimes P^B_j]/q_j$ the reduced
state of $A$ after such outcome. The quantity
\begin{equation}
I_f^{M_B}(\rho_{AB})=S_f(\rho'_{AB})-S_f(\rho_{AB})\label{dfm}
 \end{equation}
will quantify the ensuing loss of information.

We can now define the minimum of Eq.\ (\ref{dfm}) amongst all such
measurements, which will depend just on $\rho_{AB}$:
\begin{equation}
I_f^B(\rho_{AB})=\mathop{\rm Min}_{M_B} I_f^{M_B} (\rho_{AB})
 \label{Dff}\,.\end{equation}
Eq.\ (\ref{Sm}) implies $I_f^B(\rho_{AB})\geq 0$, with $I_f^B(\rho_{AB})=0$ iff
there is a complete local measurement in $B$ which leaves $\rho_{AB}$
unchanged, i.e., if $\rho_{AB}$ is already of the form (\ref{rhopab}). These
states are in general diagonal in a {\it conditional} product basis $\{|i_j
j\rangle\equiv |i_j^A\rangle\otimes|j_B\rangle\}$, where $\{|i_j^A\rangle\}$ is
the set of eigenvectors of $\rho_{A/j}$, and can be considered as {\it
partially} classical, as there is a local measurement in $B$ (but not
necessarily in $A$) which leaves them unchanged. They are the same states for
which the QD vanishes \cite{OZ.01,HV.01}. Eq.\ (\ref{Dff}) can then be
considered  a measure of the deviation of $\rho$ from such states, i.e., of
quantum correlations. One may similarly define $I_f^{A}(\rho_{AB})$ as the
minimum information loss due to a local measurement in system $A$, which may
differ from $I_f^B(\rho_{AB})$.

The states (\ref{rhopab}) are  {\it separable} \cite{RF.89}, i.e.,
convex superpositions of product states ($\rho_{AB}^s=\sum_{\alpha} q_\alpha
\rho_A^\alpha\otimes\rho_B^\alpha$, $q_\alpha>0$). Nonetheless, for a general
$\rho^s_{AB}$ the different terms $\rho_{A}^\alpha\otimes \rho_B^\alpha$ may
not commute, in contrast with (\ref{rhopab}). Hence, Eq.\ (\ref{Dff}) will be
positive not only in entangled (i.e., unseparable) states, but also in all
separable states not of the form (\ref{rhopab}), detecting those quantum
correlations emerging from the mixture of non-commuting product states.

Eq.\ (\ref{rhopab}) and concavity imply the basic bound $S_f(\rho'_{AB})\geq
\sum_j q_j S_f(\rho_{A/j})$. In addition, we also have the less trivial lower
bounds
\begin{subequations}\label{ineq1}\begin{eqnarray}
I_f^B(\rho_{AB})&\geq &S_f(\rho_A)-S_f(\rho_{AB})\,,\label{ineq2}\\
I_f^B(\rho_{AB})&\geq& S_f (\rho_B)-S_f(\rho_{AB})\,,\label{ineq3}
 \end{eqnarray}\end{subequations}
where $\rho_{A,B}={\rm Tr}_{B,A}\,\rho_{AB}$ are the local reduced states. The
r.h.s.\ in (\ref{ineq1}) is negative or zero in any separable state
\cite{NK.01,RC.02}, but can be positive in an entangled state. \\
Proof: {\it Any} separable state is more disordered globally than locally
\cite{NK.01}, as in a classical system \cite{Wh.78}:
$\rho_{AB}^s\prec\rho_{A}^s$, $\rho_{AB}^s\prec\rho^s_B$ \cite{NK.01}, or
equivalently \cite{RC.02}, $S_f(\rho^s_{AB})\geq S_f(\rho^s_{A})$,
$S_f(\rho^s_{AB})\geq S_f(\rho^s_B$) $\forall$ $f$. For the state
(\ref{rhopab}) this implies
\begin{subequations}\label{ineq}
\begin{eqnarray}
S_f(\rho_{AB}')\geq S_f(\rho'_A)=S_f(\rho_A)\,,\\
S_f(\rho_{AB}')\geq S_f(\rho'_B)\geq S_f(\rho_B)\,,
\end{eqnarray}
\end{subequations}
since $\rho_{A}'={\rm Tr}_B\,\rho'_{AB}=\sum_j q_j \rho_{A/j}=\rho_A$, while
$\rho'_B={\rm Tr}_A\,\rho'_{AB}=\sum_j q_j P_j^B$ is just the diagonal of the
actual $\rho_B$ in the basis determined by the local projectors $P_j^B$ and
hence $\rho'_B\prec \rho_B$. Eqs.\ (\ref{ineq}) lead then to Eqs.\
(\ref{ineq1}). The same inequalities (\ref{ineq1}) hold of course for
$I_f^A(\rho_{AB})$.

One may be tempted to choose as the optimal local measurement which minimizes
Eq.\ (\ref{dfm}) that based on the eigenvectors of the reduced state $\rho_B$,
in which case it will remain unchanged after measurement ($\rho'_{B}=\rho_B$).
Although this choice is optimal in the case of pure states (see \ref{IIB}) and
other relevant situations (see \ref{III}), it may not be so for a general
$\rho_{AB}$. For instance, even if local states are maximally mixed, the
optimal local measurement may not be arbitrary (see example 3 in \ref{III}).
In such a case a minor perturbation can orientate the local eigenstates along
any preferred direction, different from that where the lost information is
minimum.

\subsection{Pure states and generalized entanglement entropy\label{IIB}}
If $\rho_{AB}$ is pure ($\rho_{AB}^2=\rho_{AB}$), then
\begin{eqnarray}
I_f^B(\rho_{AB})=I_f^A(\rho_{AB})=S_f(\rho_A)=S_f(\rho_B)\,, \label{xx}
\end{eqnarray}
i.e., Eq.\ (\ref{Dff}) reduces to the generalized entropy of the subsystem
({\it generalized entanglement entropy}), quantifying the entanglement between
$A$ and $B$ according to the measure $S_f$. In the von Neumann case (\ref{S}),
Eq.\ (\ref{xx}) becomes the standard entanglement entropy \cite{EF.96}
$E_{AB}=S(\rho_A)=S(\rho_B)$, whereas in the case of the linear entropy
(\ref{SL}), Eq.\ (\ref{xx}) becomes the square of the {\it pure state
concurrence} (i.e., the tangle) \cite{Ca.03}, $C^2_{AB}=S_2(\rho_A)=
S_2(\rho_B)$.\\
Proof: For a pure state $\rho_{AB}=|\Psi_{AB}\rangle\langle\Psi_{AB}|$,
$S_f(\rho_{AB})=0$ and both $\rho_A$, $\rho_B$ have the same non-zero
eigenvalues. Eqs.\ (\ref{ineq1}) then imply $I_f^B(\rho_{AB})\geq
S_f(\rho_A)=S_f(\rho_B)$. There is also a local measurement which saturates
Eqs.\ (\ref{ineq1}): It is that determined by the Schmidt decomposition
\begin{equation}
|\Psi_{AB}\rangle=\sum_{k=1}^{n_s}
\sqrt{p_k} |k_s^A\rangle\otimes |k_s^B\rangle\,,
 \label{SD}\end{equation}
where $n_s$ is the Schmidt number and $p_k$ the non-zero eigenvalues of
$\rho_A$ or $\rho_B$ \cite{NC.00}. Choosing the local projectors in
(\ref{rhopab}) as $P_k^B=|k_s^B\rangle\langle k_s^B|$, we then obtain
\begin{equation}
 \rho'_{AB}=\sum_k p_k P_k^A\otimes P_k^B\,,\label{rap}\end{equation}
which leads to local states $\rho'_{A}=\rho_{A}=\sum_k p_k P_k^{A}$,
$\rho'_{B}=\rho_{B}=\sum_k p_k P_k^{B}$ and hence to
\begin{equation}
S_f(\rho'_{AB})=S_f(\rho_A)=S_f(\rho_B)=\sum_{k}f(p_k)\label{ko}\,,
 \end{equation}
implying Eq.\ (\ref{xx}). For pure states, entanglement can then be
considered as {\it the minimum information loss due to a local measurement,}
according to {\it any} $S_f$.

Just to verify Eq.\ (\ref{xx}), we note that for an arbitrary local measurement
defined by projectors $P_j^B=|j_B\rangle\langle j_B|$, we may rewrite Eq.\
(\ref{SD}) as
\begin{equation}|\Psi_{AB}\rangle=\sum_j
 \sqrt{q_j}|\Psi_{A/j}\rangle\otimes |j_B\rangle\,,
 \label{psiab1}\end{equation}
where $|\Psi_{A/j}\rangle=\sum_k\sqrt{p_k/q_j}\langle
j_B|k_s^B\rangle|k_s^A\rangle$ and $q_j=\sum_k p_k|\langle
j_B|k_s^B\rangle|^2$, such that $\rho_{A/j}=|\Psi_{A/j}\rangle
\langle\Psi_{A/j}|$ in (\ref{rhopab}). Hence, by concavity
$S_f(\rho'_{AB})=\sum_j f(q_j)\geq\sum_k f(p_k)$ $\forall$ $S_f$, i.e.,
$\{q_j\}\prec \{p_k\}$. Thus, for pure states, a local measurement in the basis
where $\rho_B$ is diagonal (local Schmidt basis) provides the minimum of Eq.\
(\ref{dfm}) $\forall$ $S_f$. For a maximally entangled state leading to a
maximally mixed $\rho_B$ ($p_k=1/n_B$ $\forall$ $k$) Eq.\ (\ref{dfm}) becomes
obviously independent of the choice of local basis (any choice in $B$ leads to
a corresponding basis in $A$, leaving (\ref{SD}) unchanged).

A pure state $|\Psi^I_{AB}\rangle$ can be said to be  {\it absolutely} more
entangled than another pure state $|\Psi^{II}_{AB}\rangle$ if
$S_f(\rho^I_{A})\geq S_f(\rho^{II}_A)$ $\forall$ $S_f$, i.e., if
$\rho^{I}_A\prec\rho^{II}_A$ ($\{p_k^I\}\prec\{p^{II}_k\}$). This concept has a
clear deep implication: According to the theorem of Nielsen \cite{Ni.99}, a
pure state $|\Psi^{II}_{AB}\rangle$ can be obtained from
$|\Psi^{I}_{AB}\rangle$ by local operations and classical communication (LOCC)
only if $\rho^{I}_A\prec\rho^{II}_A$, i.e., iff $|\Psi_{AB}^{I}\rangle$ is
absolutely more entangled than $|\Psi^{II}_{AB}\rangle$. This condition cannot
be ensured by a single choice of entropy, requiring the present general
measures for an entropic formulation (the exception being two-qubit or $2\times
d$ systems, where any $S_f(\rho_A)$ is a decreasing function of the largest
eigenvalue $p_1$ of $\rho_A$ and hence $S_f(\rho^I_A)\geq S_f(\rho^{II}_A)$ iff
$\rho^I_A\prec\rho^{II}_A$).

The {\it convex roof extension} \cite{Vi.00,Ca.03} of the generalized
entanglement entropy (\ref{xx}) of pure states will lead to an entanglement
measure for mixed states,
\begin{eqnarray}
E_f(\rho_{AB})=\mathop{\rm Min}_{\sum_\alpha
q_\alpha\rho_{AB}^\alpha=\rho_{AB}}\sum_\alpha q_\alpha E_f(\rho_{AB}^\alpha)
 \,, \label{Efx}\end{eqnarray}
where $q_\alpha>0$, $\rho_{AB}^\alpha=|\Psi_{AB}^\alpha
\rangle\langle\Psi_{AB}^\alpha|$ are pure states and
$E_f(\rho_{AB}^\alpha)=S_f(\rho_A^\alpha)$ is the generalized entanglement
entropy of $|\Psi_{AB}^\alpha\rangle$. Minimization is over all representations
of $\rho_{AB}$ as convex combinations of pure states. Eq.\ (\ref{Efx}) is a
non-negative quantity which clearly vanishes iff $\rho_{AB}$ is separable. It
is also an {\it entanglement monotone} \cite{Vi.00} (i.e., it cannot increase
by LOCC) since $E_f(\rho_{AB}^\alpha)$ is a concave function of
$\rho_{A}^\alpha$ invariant under local unitaries, satisfying then the
conditions of ref.\ \cite{Vi.00}. In the case of the von Neumann entropy, Eq.\
(\ref{Efx}) becomes the entanglement of formation (EOF) $E(\rho_{AB})$
\cite{EFO.96}, while in the case of the linear entropy, it leads to the mixed
state {\it tangle} $\tau(\rho_{AB})$ \cite{Ca.03,Os.05}. The general mixed
state concurrence $C(\rho_{AB})$ \cite{Ca.03} (denoted there as
$I$-concurrence) is recovered for $E_f(\rho_{AB}^\alpha)=
\sqrt{S_2(\rho_A^\alpha)}$ ($\tau=C^2$ in two qubit systems \cite{Os.05}, but
not necessarily in general).

While $I_f^B(\rho_{AB})=0$ implies $E_f(\rho_{AB})=0$ (as (\ref{rhopab}) is
separable) the converse is not true since $I_f^B(\rho_{AB})$ can be non-zero in
separable states. Nonetheless, and despite coinciding for pure states, there is
no general order relation between these two quantities for a general
$\rho_{AB}$.

\subsection{Minimum information loss by a joint local measurement}
We now consider the information loss $I_f^{M_{AB}}(\rho_{AB})$ due to a
measurement $M_{AB}$ based on {\it products}  $P^A_i\otimes P^B_j$ of one
dimensional local projectors, such that $\rho'_{AB}$ is the diagonal of
$\rho_{AB}$ in a {\it standard} product basis
$\{|ij\rangle=|i_A\rangle\otimes|j_B\rangle\}$:
\begin{equation}\rho'_{AB}=\sum_{i,j}p_{ij}P_i^A\otimes P_j^B\,,
 \label{rc}\end{equation}
where $p_{ij}=\langle ij|\rho_{AB}|ij\rangle$. Such measurement can be
considered as a subsequent local measurement in $A$ after a measurement in $B$
(if the results are of course unknown),  implying $I_f^{M_{AB}}(\rho_{AB})\geq
I_f^{M_B}(\rho_{AB})$, where $M_B=\{P_j^B\}$ is the measurement in $B$. The
ensuing minimum
\begin{equation}
I_f^{AB}(\rho_{AB})=\mathop{\rm Min}_{M_{AB}}I_f^{M_{AB}}(\rho_{AB})\,,
 \label{dfab}\end{equation}
will then satisfy in general
\begin{equation} I_f^{AB}(\rho_{AB})\geq I_f^B(\rho_{AB})\,,
 \label{ineq0}\end{equation}
with $I_f^{AB}(\rho_{AB})=0$ if and only if $\rho_{AB}$ is of the form
(\ref{rc}). The state (\ref{rc}) represents a {\it classically correlated
state} \cite{Lu2.08,LL.08}. Fur such states there is a local measurement in $A$
as well as in $B$ which leaves the state unchanged, being equivalent in this
product basis to a classical system described by a joint probability
distribution $p_{ij}$. Eq.\ (\ref{dfab}) is then a measure of all quantum-like
correlations. The states (\ref{rc}) are of course a particular case of
(\ref{rhopab}), i.e., that where all $\rho_{A/j}$ are mutually commuting.
Product states $\rho_A\otimes \rho_B$ are in turn a particular case of
(\ref{rc}) ($p_{ij}=p^A_ip^B_j$ $\forall$ $i.j$) and correspond to $\rho_{A/j}$
independent of $j$ in (\ref{rhopab}).

In the case of pure states we obtain, however,
\begin{equation}
I_f^{AB}(\rho_{AB})=I_f^B(\rho_{AB})=S_f(\rho_A)=S_f(\rho_B)\,,
\label{xxx}
\end{equation}
since the state (\ref{rap}) is already of the form (\ref{rc}), being left
unchanged  by a measurement based on the Schmidt basis projectors
$P_{k'}^A\otimes P_k^B$. Pure state entanglement can then be also seen as {\it
the minimum information loss due to a joint local measurement}.

For an arbitrary product measurement on a pure state, the expansion
\begin{equation}|\Psi_{AB}\rangle=\sum_{i,j}c_{ij}|i_A\rangle\otimes
 |j_B\rangle\,, \label{psiab2}\end{equation}
with $c_{ij}=\sum_{k}\sqrt{p_k}\langle i_A|k_s^A\rangle\langle
j_B|k_s^B\rangle$, leads to $p_{ij}=|c_{ij}^2|$ in (\ref{rc}). Eqs.\
(\ref{rc})--(\ref{xxx}) then imply $S_f(\rho'_{AB})=\sum_{i,j} f(|c^2_{ij}|)\geq
\sum_k f(p_k)$ $\forall$ $S_f$. Since
$I_f^{M_{AB}}(\rho_{AB})\geq I_f^{M_{B}}(\rho_{AB})\geq I_f^B(\rho_{AB})$,
Eqs.\ (\ref{psiab1}), (\ref{xxx}) and (\ref{psiab2}) lead to
\begin{equation}\{|c_{ij}^2|\}\prec \{q_j\}\prec \{p_k\}\,.\label{may}
 \end{equation}
The first relation is apparent as $q_j=\sum_i |c_{ij}^2|$ is just the marginal
of the joint distribution $|c_{ij}^2|$. The state (\ref{rap}) can then be
rigorously regarded  as {\it the closest classical state} to the pure state
$\rho_{AB}$, since it provides the lowest information loss among {\it all}
local or joint local measurements for {\it any} $S_f$. Pure states have
therefore an associated {\it least mixed classical state}, such that the state
obtained after any local measurement is always majorized by it.

Let us finally mention that it is also feasible to consider more general
product measurements $M_{A/B}$ based on conditional product projectors
$P_{i_j}^A\otimes P_j^B$, leading to a $\rho'_{AB}$ diagonal in a conditional
product basis,
\begin{equation}
\rho'_{AB}=\sum_{i,j} p_{ij}P_{i_j}^A\otimes P_j^B\label{cond}
\end{equation}
where $p_{ij}=\langle i_j j|\rho_{AB}|i_j j\rangle$. The ensuing information
loss will satisfy again $I_f^{M_{A/B}}(\rho_{AB})\geq I_f^{M_B}(\rho_{AB})$, as
(\ref{cond}) can still be considered as the diagonal of (\ref{rhopab}) in a
conditional product basis $\{|i_j j\rangle\}$, where the
$\{|i_j^A\rangle\}$ are  not necessarily the eigenvectors of $\rho_{A/j}$.
However, if chosen as the latter, we have
$I_f^{M_{A/B}}(\rho_{AB})=I_f^{M_B}(\rho_{AB})$ and hence,
\begin{equation}
I_f^{A/B}(\rho_{AB})=\mathop{\rm Min}_{M_{A/B}}I_f^{M_{A/B}}
(\rho_{AB})=I_f^B(\rho_{AB})\,,
\end{equation}
as (\ref{rhopab}) remains unchanged under a measurement in the optimum
conditional product basis formed by the eigenvectors of the $\rho_{A/j}$ times
the states $|j_B\rangle$.

\subsection{Von Neumann based measures}
If $S_f(\rho)$ is chosen as the von Neumann entropy (\ref{S}), Eq.\ (\ref{dfm})
becomes (see Eq.\ (\ref{DsB}))
\begin{equation} I^{M_B}(\rho_{AB})=S(\rho'_{AB})-S(\rho_{AB})
=S(\rho_{AB}||\rho'_{AB})\,.
\end{equation}
The ensuing minimum $I^B(\rho_{AB})$ is also the {\it minimum} relative entropy
between $\rho_{AB}$ and any state $\rho^d_{AB}$ diagonal in a standard or
conditional product basis:
\begin{equation}
I^{B}(\rho_{AB})=\mathop{\rm Min}_{M_B} I^{M_B}(\rho_{AB})=
\mathop{\rm Min}_{\rho^d_{AB}} S(\rho_{AB}||\rho^d_{AB})\,,\label{IB}
\end{equation}
where $\rho^d_{AB}$ denotes a state of the general form (\ref{rhopab}) with
{\it both} the local projectors $P_j^B=|j_B\rangle\langle j_B|$ as well as the
probabilities $q_j$ and states $\rho_{A/j}$ being arbitrary.

Proof: For a given choice of conditional product basis, the minimum relative
entropy is obtained when  $\rho^d_{AB}$ has the same diagonal elements as
$\rho_{AB}$ in that basis (as $-\sum_i p_i\log_2 q_i$ is minimized for
$q_i=p_i$). Hence, $S(\rho_{AB}||\rho^d_{AB})\geq
S(\rho_{AB}||\rho'_{AB})=I^{M_{A/B}}(\rho_{AB})\geq I^B(\rho_{AB})$, where
$\rho'_{AB}$ denotes here the post-measurement state (\ref{cond}) in that
basis.

The same property holds for $I^{AB}(\rho_{AB})$ if $\rho^d_{AB}$ is restricted
to states diagonal in a standard product basis:
\begin{equation}
I^{AB}(\rho_{AB})=\mathop{\rm Min}_{M_{AB}} I^{M_{AB}}(\rho_{AB})= \mathop{\rm
Min}_{\rho^d_{AB}} S(\rho_{AB}||\rho^d_{AB})\,, \label{IABr}
\end{equation}
where $\rho^d_{AB}$ is here of the form (\ref{rc}) with $p_{ij}$ arbitrary.
Eq.\ (\ref{IABr}) is precisely the bipartite version of the quantity $D$
introduced in \cite{Ve.10} as a measure of quantum correlations for composite
systems.

The quantity (\ref{IB}) is also closely related to the {\it quantum discord}
\cite{OZ.01,HV.01,DSC.08}, which can be written in the present notation as
$D^B(\rho_{AB})=\mathop{\rm Min}_{M_B}D^{M_B}(\rho_{AB})$, with
\begin{eqnarray}
D^{M_B}(\rho_{AB})&=&S(\rho'_{AB})-S(\rho'_B)-[S(\rho_{AB})-S(\rho_{B})]
 \label{Dmb}\,,\\&=&I^{M_B}(\rho_{AB})-I^{M_B}(\rho_B)\,,\end{eqnarray}
where $\rho'_{AB}$ is the measured state (\ref{rhopab}) and  $\rho'_B$,
$\rho_B$ the reduced states after and before the measurement. Thus,
$D^B(\rho_{AB})\leq I^B(\rho_{AB})$. They will coincide when the optimal local
measurement is the same for both (\ref{dfm}) and (\ref{Dmb}) and corresponds to
the basis where $\rho_B$ is diagonal, such that $\rho'_B=\rho_B$
($I^{M_B}(\rho_B)=0$). This coincidence takes place, for instance, whenever
$\rho_B$ is maximally mixed (as in this case $\rho'_B=\rho_B$ for any choice of
local basis). Both $D^B(\rho_{AB})$ and $I^B(\rho_{AB})$ also vanish for the
same type of states (i.e., those of the form (\ref{rhopab})) and both reduce to
the standard entanglement entropy $E_{AB}=S(\rho_A)$ for pure states (although
Eq.\ (\ref{Dff}) requires a measurement in the local Schmidt basis whereas
(\ref{Dmb}) becomes independent of the choice of local basis, as $\rho_{A/j}$
is pure and hence $S(\rho'_{AB})=S(\rho'_B)$ for any local measurement). A
direct generalization of (\ref{Dmb}) to a general entropy $S_f(\rho)$ is no
longer positive for a general concave $f$, since the positivity of (\ref{Dmb})
relies on the concavity of the {\it conditional} von Neumann entropy
$S(A|B)=S(\rho_{AB})-S(\rho_B)$ \cite{Wh.78}, which does not hold for a general
$S_f$.

Minimum distances between $\rho_{AB}$ and classical states of the form
(\ref{rc}) were also considered in \cite{Lu2.08}, where the attention was
focused on the decrease $Q$ of the mutual information
$S(\rho_A)+S(\rho_B)-S(\rho_{AB})$ after a measurement $M_{AB}$ in the product
basis formed by the eigenstates of $\rho_A$ and $\rho_B$. Such quantity
coincides with present $I^{M_{AB}}(\rho_{AB})$ for this choice of basis as
$\rho_A$ and $\rho_B$ remain unchanged. Nonetheless, for a general $\rho_{AB}$
the minimum (\ref{IABr}) may be attained at a different basis.

\subsection{Quadratic measure}
If $S_f(\rho)$ is chosen as the linear entropy (\ref{SL}), Eq.\ (\ref{dfm})
becomes (see Eq.\ (\ref{D2b}))
 \begin{equation}
I_2^{M_B}(\rho_{AB})=2{\rm Tr}(\rho_{AB}^2-{\rho'}_{AB}^2)=
2||\rho_{AB}-\rho'_{AB}||^2\,, \label{DLm}
 \end{equation}
where $||\rho_{AB}-\rho'_{AB}||^2=\sum\limits_{j\neq j',i,k}|\langle
ij|\rho_{AB}|kj'\rangle|^2$ is just the squared norm of the off-diagonal
elements lost after the local measurement. It therefore provides the simplest
measure of the information loss. Its minimum is the {\it minimum} squared
Hilbert-Schmidt distance between $\rho_{AB}$ and {\it any} state $\rho^d_{AB}$
diagonal in a general product basis:
 \begin{equation}
I_2^B(\rho_{AB})=\mathop{\rm Min}_{M_B}\,I^{M_B}_2(\rho_{AB})=
\mathop{\rm Min}_{\rho_{AB}^d}||\rho_{AB}-\rho^d_{AB}||^2\,, \label{I2B}
 \end{equation}
where the last minimization is again over all states of the form
(\ref{rhopab}), with $P_j^B$, $q_j$ and $\rho_{A/j}$ arbitrary.
\\
Proof: For a general product basis, $||\rho_{AB}-\rho_{AB}^d||^2=
||\rho_{AB}-\rho'_{AB}||^2+ ||\rho'_{AB}-\rho^d_{AB}||^2$, where $\rho'_{AB}$
is again the diagonal of $\rho_{AB}$ in this basis. Hence, the optimum choice
in this basis is $\rho_{AB}^d=\rho'_{AB}$, whence
$||\rho_{AB}-\rho_{AB}^d||^2\geq
||\rho_{AB}-\rho_{AB}'||^2=I_2^{M_{A/B}}(\rho_{AB})\geq I_2^B(\rho_{AB})$.
Actually, we could also extend the last minimization in (\ref{I2B}) to all
operators $O^d_{AB}$ diagonal in a general product basis.

The same property holds for $I_2^{AB}(\rho_{AB})$ if $\rho^d_{AB}$ is
restricted to states diagonal in a standard product basis:
\begin{equation}
I_2^{AB}(\rho_{AB})=\mathop{\rm Min}_{M_{AB}} I_2^{M_{AB}}(\rho_{AB})=
\mathop{\rm Min}_{\rho^d_{AB}} ||\rho_{AB}-\rho^d_{AB}||^2\,,\label{I2AB}
\end{equation}
where $\rho^d_{AB}$ is here of the general form (\ref{rc}). Note that
$I_2^{M_{AB}}(\rho_{AB})=I_2^{M_B}(\rho_{AB})+\sum_{j,\, i\neq k}|\langle
ij|\rho_{AB}|kj\rangle|^2$ is just the squared norm of all off-diagonal
elements.

In the case of pure states, Eqs.\ (\ref{I2B}) and (\ref{I2AB}) reduce to the
pure state concurrence \cite{Ca.03} $C^2_{AB}=S_2(\rho_A)$.

\section{Examples \label{III}}
We will now evaluate the general measures (\ref{Dff}) and (\ref{dfab}) for any
$S_f$ in a few simple relevant examples.
\subsection{Mixture of a general pure state with the maximally mixed state}
For a convex mixture of $|\Psi_{AB}\rangle=\sum\limits_{k=1}^{n_s}
\sqrt{p_k}|k_s^Ak_s^B\rangle$ (Eq.\ (\ref{SD})) with the maximally mixed state,
i.e.,
\begin{equation}
\rho_{AB}(x)=x|\Psi_{AB}\rangle\langle\Psi_{AB}|+
{\textstyle\frac{1-x}{n}}I_A\otimes I_B\,,\label{rx}
 \end{equation}
where $x\in[0,1]$ and $n=n_A n_B$, the minimum $I_f^B(x)\equiv
I_f^B[\rho_{AB}(x)]$ corresponds again to a measurement in the local Schmidt
basis for $|\Psi_{AB}\rangle$ and is given by
\begin{equation}
I_f^B(x)=
\sum_{k=1}^{n_s}[f(xp_k+{\textstyle\frac{1-x}{n}})-f(\delta_{k1}x+
{\textstyle\frac{1-x}{n}})]
\,,\label{Dfx}
 \end{equation}
with $I_f^A(x)=I_f^{AB}(x)=I_f^B(x)$. Eq.\ (\ref{Dfx}) is a {\it strictly
increasing function of} $x$ $\forall$ $S_f$ if $n_s\geq 2$ (i.e., if
$|\Psi_{AB}\rangle$ is entangled), implying $I_f^B(x)>0$ $\forall$ $x\in(0,1]$.

Proof: After a local measurement in the basis $\{|k_s^B\rangle\}$, the joint
state becomes
\begin{equation}
\rho'_{AB}(x)=x\sum_{k=1}^{n_s}p_k P_k^A\otimes P_k^B+{\textstyle\frac{1-x}{n}}
I_A\otimes I_B\,,\label{rhopx}
 \end{equation}
which is diagonal in the Schmidt basis $\{|{k'}_s^A\rangle\otimes
|k_s^B\rangle\}$ with diagonal elements $p_{k'k}=\delta_{k'k}x
p_k+\frac{1-x}{n}$. For any other complete local measurement, $\rho'_{AB}$ will
be diagonal in a basis $\{|i_j^A\rangle\otimes|j_B\rangle\}$, where we set
$|j_j^A\rangle=|\Psi_{A/j}\rangle$ (Eq.\ (\ref{psiab1})), with diagonal
elements $p'_{ij}=\delta_{ij}xq_j+\frac{1-x}{n}$. The latter are always
majorized by $p_{kk'}$ ($\{p'_{ij'}\}\prec\{p_{kk'}\}$) since $\{q_j\}\prec
\{p_k\}$ (Eq.\ (\ref{may})) and $x\geq 0$. Hence, $S_f(\rho'_{AB})$ is minimum
for a measurement in the basis $\{|k_s^B\rangle\}$, which leads to Eq.\
(\ref{Dfx}). Moreover, $I_f^{AB}(x)=I_f^{A}(x)=I_f^B(x)$ since (\ref{rhopx}) is
diagonal in a standard product basis.

Eq.\ (\ref{rhopx}) is again the {\it closest} classical state to (\ref{rx}),
majorizing {\it any} other state obtained after a local or product measurement.

To verify the monotonicity, we note that
\begin{eqnarray}
\frac{d I_f^B}{dx}&=&\sum_{k=1}^{n_s}[(p_k-{\textstyle\frac{1}{n}})
 f'(p_k^x)-(\delta_{k1}-{\textstyle\frac{1}{n}})f'(\lambda_k^x)]\nonumber\\
&\geq&({\textstyle\frac{n'_s-1}{n}}+\!\!\sum_{p_k<1/n}p_k)
[f'(\lambda_2^x)-f'(\lambda_1^x)]\geq 0\,,
 \end{eqnarray}
since $\lambda_2^x=\frac{1-x}{n}\leq p_k^x\leq \lambda_1^x=x+\frac{1-x}{n}$ and
hence $f'(\lambda_2^x)\geq f'(p_k^x)\geq f'(\lambda_1^x)$, where
$p_k^x=xp_k+\frac{1-x}{n}$ and $n'_s\geq 1$ is the number of Schmidt
probabilities $p_k$ not less than $1/n$. Eq.\ (\ref{Dfx}) is then strictly
increasing if $f$ is strictly concave and $n_s\geq 2$, implying $I_f^B(x)=0$
only if $x=0$ or $n_s=1$.

A series expansion of (\ref{Dfx}) around $x=0$ shows that
\begin{equation}
I_f^B(x)=-\half x^2 f''({\textstyle\frac{1}{n}})
 (1-\sum_k p_k^2)+O(x^3)\,,\label{x2g}
 \end{equation}
in agreement with Eq.\ (\ref{quad}), indicating {\it a universal quadratic
increase} of $I_f^B(x)$ for small $x$  ($f''(1/n)<0$). For the quadratic measure
(\ref{DLm}) we obtain in fact a simple quadratic dependence $\forall$
$x\in[0,1]$:
\begin{equation}
I_2^B(x)=x^2 I_2^B(1)=2x^2(1-\sum_k p_k^2)\label{q2}\,.
\end{equation}
Hence, for $|\Psi_{AB}\rangle$ entangled, $I_f^B(x)>0$ as soon as the mixture
(\ref{rx}) departs from the maximally mixed state. In contrast, any
entanglement measure, like the monotones (\ref{Efx}) or the negativity
\cite{VW.02}, requires a {\it finite} threshold value $x_c>0$, since Eq.\
(\ref{rx}) is {\it separable} for small $x$: Any bipartite state $\rho$ is
separable if ${\rm Tr}(\rho-I_n/n)^2\leq \frac{1}{n(n-1)}$ \cite{GB.02}, which
ensures here separability for $x\leq \frac{1}{n-1}\leq x_c$ ($n\geq 4$). In the
maximally entangled case $p_k=1/d$, with $n_A=n_B=d$, (\ref{rx}) is in fact
separable iff $x\leq 1/(d+1)$ \cite{Ca.03,HH.99}. In general, the negativity
will be positive for $x>x_c=\frac{1}{1+n \sqrt{p_1 p_2}}$, sorting the $p_k$ in
decreasing order.

Let us finally notice that given two pure states $|\Psi^{I}_{AB}\rangle$ and
$|\Psi^{II}_{AB}\rangle$, the ensuing mixtures (\ref{rx}) will satisfy, at
fixed $x\in(0,1]$, $I_f^{B I}(x)\geq I_f^{B II}(x)$ $\forall$ $S_f$ iff
$|\Psi^I_{AB}\rangle$ is {\it absolutely} more entangled than
$|\Psi_{AB}^{II}\rangle$ ($\{p^{I}_k\}\prec\{p^{II}_k\}$). This is apparent as
$S_f(\rho^{I}_{AB}(x))=S_f(\rho^{II}_{AB}(x))$ whereas
${\rho'}_{AB}^I(x)\prec{\rho'}_{AB}^{II}(x)$ iff $\{p^{I}_k\}\prec
\{p^{II}_k\}$ (Eq.\ (\ref{rhopx})), in which case $S_f({\rho'}_{AB}^I(x))\geq
S_f({\rho'}_{AB}^{II}(x))$.

\subsection{Two-qubit case}
Let us now explicitly consider the mixture (\ref{rx}) in the two-qubit case,
where $|\Psi_{AB}\rangle$ can be always written as
\begin{equation}
|\Psi_{AB}\rangle=\sqrt{p}\,|00\rangle+\sqrt{1-p}\,|11\rangle\,,\label{psab}
\end{equation}
with $|ij\rangle\equiv |i_s^A\rangle\otimes|j_s^B\rangle$ and $p\in[0,1]$. For
a local spin measurement along an axis forming an angle $\theta$ with the $z$
axis, it is easy to show that the information loss is
\begin{equation}
I_f^{\theta_B}(x)=\sum_{\nu=\pm}[{\textstyle
f(\frac{1+x(1+2\nu \cos\theta(2p-1))}{4})-f(\frac{1+2\nu x}{4})}]\label{dxt}\,.
 \end{equation}
It is verified that for $p\neq 1/2$, $I_f^{\theta_B}(x)$ is minimum for
$\theta=0$, i.e., for a measurement in the local Schmidt basis for
$|\Psi_{AB}\rangle$ (as $\rho'(\theta)\prec\rho'(0)$), while for $p=1/2$ (Bell
state)  $I_f^{\theta_B}(x)$ is $\theta$-independent, as the local Schmidt basis
becomes arbitrary.  The minimum  becomes then
\begin{equation}
I_f^B(x)={\textstyle
f(\frac{1+x(4p-1)}{4})+f(\frac{1+x(3-4p)}{4})-f(\frac{1+3x}{4})
-f(\frac{1-x}{4})} \label{dx}
 \end{equation}
(Eq.\ (\ref{Dfx})), being a strictly increasing function of $x$ if $f$ is
strictly concave and $p\in(0,1)$ (if $p=0$ or $1$, $|\Psi_{AB}\rangle$ is
separable and  $I_f^B(x)=0$ $\forall$ $x$). It is also a decreasing function of
$p$ for $p\in[\half,1]$ at fixed $x$.

\begin{figure}
\vspace*{-0.25cm}

\centerline{\scalebox{.85}{\includegraphics{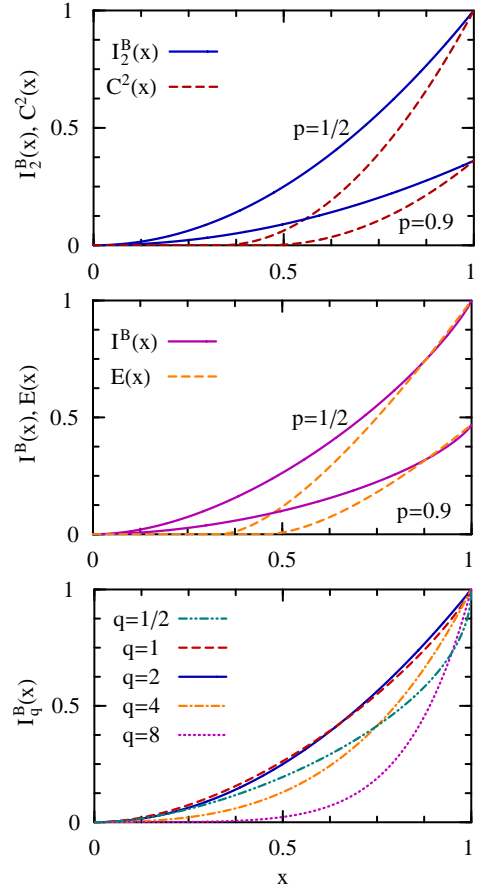}}}
 \vspace*{-0.75cm}

\caption{Measures of quantum correlations and entanglement for the mixture
(\ref{rx}) of the pure state (\ref{psab}) with the maximally mixed state, for
$p=1/2$ and $p=0.9$. Top: The quadratic measure $I_2^B$ (Eqs.\ (\ref{I2B}),
(\ref{d2x})) and the squared concurrence $C^2$ (Eq.\ (\ref{Cx})), satisfying
$I_2^B(x)>C^2(x)$ for $x\in(0,1)$. Center: The von Neumann based measure $I^B$
(Eqs.\  (\ref{IB}), (\ref{dx}) for $f(p)=-p\log_2 p$) and the entanglement of
formation $E$, again coincident for $x=1$ but exhibiting no fixed order
relation for $x\in(0,1)$. Bottom: Behavior of $I^B_q$ (Eq.\ (\ref{dx}) for
$f(p)=\frac{p-p^q}{1-2^{1-q}}$) for different $q$ and $p=1/2$. $I_1^B=I^B$ is
the von Neumann measure while $I_2^B$ the quadratic measure.}
 \label{f1}
\end{figure}
 \vspace*{0.cm}

In particular, for $S_f(\rho)=S_2(\rho)$, Eq.\ (\ref{dx}) becomes
\begin{eqnarray}
I^B_{2}(x)&=&4x^2p(1-p)\label{d2x}\,.
\end{eqnarray}
We may compare (\ref{d2x}) with the corresponding entanglement monotone
(\ref{Efx}) (the tangle), which coincides here with the squared concurrence
\cite{WW.98,Ca.03} $C^2(x)$ of $\rho_{AB}(x)$. For a general two-qubit mixed
state the concurrence can be calculated as \cite{WW.98} $C={\rm
Max}[2\lambda_M-{\rm Tr}R,0]$, where $\lambda_M$ is the largest eigenvalue of
$R=\sqrt{\rho_{AB}^{1/2} \tilde{\rho}_{AB}\rho^{1/2}_{AB}}$, with
$\tilde{\rho}_{AB}=\sigma_y^A\otimes \sigma_y^B\rho_{AB}^*\sigma_y^A\otimes
\sigma_y^B$. This leads here to
\begin{equation}
C(x)={\rm Max}[2x\sqrt{p(1-p)}-{\textstyle\frac{1-x}{2}},0]\,,\label{Cx}
\end{equation}
which vanishes for $x\leq x_c=\frac{1}{1+4\sqrt{p(1-p)}}$. It is then verified
that for the present mixture,
\[I_2^B(x)\geq C^2(x)\,,\]
$\forall$ $p$, $x$, with $I_2^B(x)=C^2(x)$ just for $x=0$ or $x=1$ if
$p\in(0,1)$, as seen  in the top panel of Fig.\ \ref{f1} (such inequality does
not hold for any two-qubit mixed state).

In contrast, the von Neumann based measure $I^B(x)$ (Eq.\ (\ref{IB})) is not an
upper bound to the EOF $E(x)$ of $\rho_{AB}(x)$, as seen in the central panel,
even though they both coincide for $x=1$ $\forall$ $p$. For any two qubit
state, $E$ can be evaluated in terms of the concurrence $C$ as \cite{WW.98}
\begin{equation}
E=\sum_{\nu=\pm}f({\textstyle\frac{1+\nu\sqrt{1-C^2}}{2}})\,,\label{EAB}
\end{equation}
for $f(p)=-p\log_2 p$, which is just the relation between $S_f(\rho)$ and
$S_2(\rho)=C^2$ for a single qubit state $\rho$. Hence, for $x$ close to $1$,
$E(x)-I^B(x)\approx-\frac{1-x}{4}\log_2(1-x)>0$, as $E(x)$ decreases linearly
whereas $I^B(x)$ decreases logarithmically. Notice that $I^B(x)$ coincides with
the QD $\forall$ $p,x$, as Eq.\ (\ref{Dmb}) is also minimized by a measurement
along the $z$ axis ($\theta=0$), in which case $\rho'_B=\rho_B$.

The bottom panel depicts the behavior of Eq.\ (\ref{dx}) for the Tsallis case
$f(p)=f_q(p)\equiv\frac{p-p^q}{1-2^{1-q}}$. As $q$ increases above 2,
$I_q^B(x)$ becomes less sensitive to weak quantum correlations (as $f_q''(1/n)$
in (\ref{x2g}) becomes small), resembling the behavior of the entanglement
measures.

One may here ask if it is also possible to employ Eq.\ (\ref{EAB}) with a
general $f$ for evaluating the corresponding {\it generalized} EOF (\ref{Efx}).
According to the arguments of \cite{WW.98} and \cite{Ki.10}, this is feasible
provided Eq.\ (\ref{EAB}), which is a strictly increasing function of $C$
$\forall$ concave $f$, is also {\it convex}. In the Tsallis case $f(p)=f_q(p)$,
this allows the applicability of (\ref{EAB}) for $\frac{5-\sqrt{13}}{2}<q<
\frac{5+\sqrt{13}}{2}$ (as obtained from the condition $E''(C)\geq 0$ $\forall$
$C\in[0,1]$), i.e., $0.7\alt q\alt 4.3$, in agreement with the numerical
results of \cite{Ki.10}. Denoting the ensuing quantity as $E_q(x)$, we then
obtain, for the present normalization,
 \begin{equation} E_{2}=E_{3}=C^2\,,\label{E23}\end{equation}
as for any single qubit state $\rho$, $S_2(\rho)=S_3(\rho)=4\,{\rm det}(\rho)$.

The inequality $I_q^B(x)\geq E_q(x)$ $\forall$ $x\in[0,1]$ will then hold in a
certain {\it finite} interval around $q=2$, namely $1.27\alt q\alt 3.5$ for
$p=1/2$ and $1.3\alt q\alt 4.3$ for $p=0.9$. These boundaries are actually
determined by the slope condition ${I_q^B}'(1)<E'_q(1)$. For instance, for
$p=1/2$ and a general entropic $f$ such that (\ref{EAB}) is convex, we have
\begin{eqnarray}
I_f^B(x)&\approx& 1-{\textstyle\frac{1}{4}[f'(0)+2f'(\frac{1}{2})
-3f'(1)](1-x)}\,,\\
 E_f(x)&\approx & 1+{\textstyle\frac{3}{4}
 f''(\frac{1}{2})(1-x)}\,,\end{eqnarray}
for $x\rightarrow 1$, such that $I_f^B(x)>E_f(x)$ in this limit iff
$f'(0)+2f'(1/2)-3f'(1)< -3f''(1/2)$. This leaves out the von Neumann entropy
($f'(0)\rightarrow\infty$) as well as all $q<1$ in the Tsallis case, leading in
the latter to the previous interval $1.27\alt q\alt 3.5$.
\subsection{Decoherence of a Bell state}
Let us now consider the state
\begin{eqnarray}
\rho_{AB}(z)&=&\half[|00\rangle\langle 00|+|11\rangle\langle 11|+
z(|00\rangle\langle 11|+|11\rangle\langle 00|)]\nonumber\\
&=&{\textstyle \frac{1+z}{2}}|\Psi_+\rangle\langle\Psi_+|
+{\textstyle\frac{1-z}{2}}|\Psi_-\rangle\langle\Psi_-|\,,
 \label{st2}\end{eqnarray}
where $|z|\leq 1$ and $|\Psi_\pm\rangle=\frac{|00\rangle\pm
|11\rangle}{\sqrt{2}}$. It corresponds to the partial decoherence of
$|\Psi_\pm\rangle$ and can be also seen as a mixture of these two Bell states.
Even though the reduced states $\rho_A$ and $\rho_B$ are maximally mixed
$\forall$ $z$, a local spin measurement along an axis forming an angle $\theta$
with the $z$ axis leads to a post-measurement state $\rho'(\theta)$ with
two-fold degenerate eigenvalues $\frac{1\pm\sqrt{1-\sin^2\theta(1-z^2)}}{4}$
and hence, to a $\theta$-{\it dependent} information loss
\begin{equation}
I_f^{\theta_B}(z)=
\sum_{\nu=\pm}[{\textstyle 2f(\frac{1+\nu\sqrt{1-
\sin^2\theta(1-z^2)}}{4})-f(\frac{1+\nu z}{2})}]\,.
 \end{equation}

\begin{figure}[t]
\vspace*{-.4cm}

\centerline{\scalebox{.85}{\includegraphics{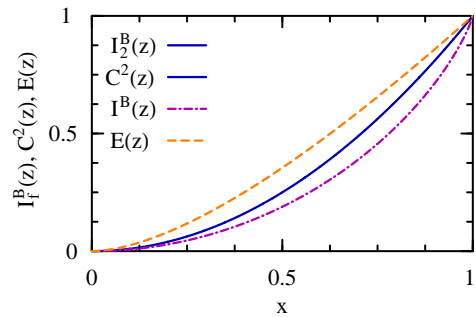}}}
 \vspace*{-.6cm}

\caption{Same details as fig.\ \ref{f1} for the state (\ref{st2}). Here
$I_2^B(z)=C^2(z)$ whereas $E(z)\geq I^B(z)$ for $z\in(0,1)$.}
 \label{f2}
\end{figure}
 \vspace*{0cm}

Its minimum for any $z\in(-1,1)$ and concave $f$ corresponds again to
$\theta=0$, as $\rho'(\theta)\prec\rho'(0)=\half(|00\rangle\langle
00|+|11\rangle\langle 11|)$ $\forall$ $\theta$. We then obtain, setting again
$2f(1/2)=1$,
\begin{equation} {\textstyle I_f^B(z)=
1-f(\frac{1+z}{2})-f(\frac{1-z}{2})}\,,
\label{Ifz}
\end{equation}
with $I_f^{AB}(z)=I_f^A(z)=I_f^B(z)$ as $\rho(0)$ is diagonal in a standard
product basis. Hence, $I_f^B(z)>0$ if $z\neq 0$, with $I_f^B(z)=
-\frac{1}{4}f''(\frac{1}{2}) z^2+O(z^3)$ for $z\rightarrow 0$. Moreover,
$I_f^B(z)$ is an increasing (${I_f^B}'(z)>0$) convex (${I_f^B}''(z)>0$)
function of $z$ $\forall$ $S_f$.

In the case of the linear entropy, Eq.\ (\ref{Ifz}) becomes
\begin{equation} I_2^B(z)=z^2=C^2(z) \,, \end{equation}
where $C(z)=|z|$ is the concurrence of (\ref{st2}). Thus, here $I_2^B(z)$ and
$E_2(z)$ {\it coincide exactly} $\forall$ $z\in[0,1]$. In contrast, the von
Neumann measure $I^B(z)$ is {\it smaller} than the EOF $E(z)=\sum_{\nu=\pm}
f(\frac{1+\nu\sqrt{1-z^2}}{2})$ ($f(p)=-p\log_2 p$) $\forall$ $z\in(0,1)$
(Fig.\ \ref{f2}). For small $z$ we have in particular $E(z)\approx
-\frac{1}{2}z^2\log_2 z^2>I^B(z)\approx \half z^2/\ln 2$. Again, $I^B(z)$
coincides here with the QD as $\rho_B$ is maximally mixed.

Let us finally remark that Eqs.\ (\ref{E23}) and (\ref{Ifz}) also imply
 \[I_3^B(z)=z^2=E_3(z)\,.\]
It can then be seen that for $2<q<3$, $I_q^B(z)>E_q(z)$ $\forall$ $z\in(0,1)$
(although the difference is small) whereas for $q<2$ or $q>3$ (within the
limits allowed by the validity of (\ref{EAB})) $I_q^B(z)<E_q(z)$ $\forall$
$z\in(0,1)$. These intervals can be corroborated from the expansions for
$z\rightarrow 0$ and $z\rightarrow 1$,
\begin{eqnarray}
I_f^B(z)-E_f(z)&=& {\textstyle\frac{1}{4}[-f''(\frac{1}{2})-f'(0)+f'(1)]z^2+
O(z^3)}\,,\nonumber\\
&= & {\textstyle\frac{1}{4}[-f''(\frac{1}{2})-f'(0)+f'(1)](1-z)+O(1-z)^2}
 \nonumber \end{eqnarray}
which imply $I_f^B(z)>E_f(z)$ in these limits iff $f'(0)-f'(1)<-f''(1/2)$,
leading to $2<q<3$ in Tsallis case.

\section{Conclusion \label{IIII}}
We have constructed a general entropic measure of quantum correlations
$I_f^B(\rho_{AB})$, which represents the minimum loss of information, according
to the entropy $S_f$, due to a local projective measurement. Its basic
properties are similar to those of the quantum discord, vanishing for the same
partially classical states (\ref{rhopab}) and coinciding with the corresponding
generalized entanglement entropy in the case of pure states. Its positivity
relies, however, entirely on the majorization relations fulfilled by the
post-measurement state, being hence applicable with general entropic forms
based on arbitrary concave functions. In particular, for the linear entropy it
leads to a quadratic measure $I_2^B(\rho_{AB})$ which is particularly simple to
evaluate and can be directly interpreted as minimum squared distance, yet
providing the same qualitative information as other measures. The minimum loss
of information due to a joint local measurement $I_f^{AB}(\rho_{AB})$, has also
been discussed, and shown to coincide with $I_f^B(\rho_{AB})$ in some important
situations, vanishing just for the classically correlated states (\ref{rc}).

While there is no general order relation between these quantities and the
associated entanglement monotones (\ref{Efx}), the use of generalized entropies
allows at least to find such a relation in some particular cases: The quadratic
measure $I_2^B(\rho_{AB})$ provides for instance an upper bound to the squared
concurrence of the two-qubit states (\ref{rx})--(\ref{psab}) (unlike the von
Neumann based measures) and coincides with it in the mixture (\ref{st2}).
Moreover, generalized entropies such as $S_q(\rho)$ allow to find in these
previous cases an interval of $q$ values where an order relationship holds,
which requires a delicate balance between the derivatives of $f$ at different
points.

Let us finally mention that some general concepts emerge naturally from the
present formalism, like that of absolutely more entangled and in particular
that of the {\it least mixed} classically correlated state that can be
associated with certain states, such as pure states or the mixtures (\ref{rx})
or (\ref{st2}). This state majorizes any other state obtained after a local
measurement, thus minimizing the entropy increase (\ref{dfm}) or (\ref{dfab})
for {\it any} choice of entropy $S_f$. It allows for an unambiguous
identification of the least perturbing local measurement.

The authors acknowledge support of CIC (RR) and CONICET (LC, NC) of Argentina.

\end{document}